\documentclass[preprint,12pt]{elsarticle}

\usepackage{amssymb}
\usepackage{amsmath}

\usepackage[colorinlistoftodos]{todonotes} 
\usepackage{amsmath,amssymb,amsfonts}
\usepackage{algorithmic}
\usepackage{graphicx}
\graphicspath{{figures/}}
\usepackage{textcomp}
\usepackage{xcolor}
\usepackage{comment}
\usepackage{hyperref}
\usepackage{longtable}
\usepackage{blindtext}
\usepackage{rotating}
\usepackage{multirow}
\usepackage{enumitem}
\usepackage{tcolorbox}

\journal{Journal of Systems and Software}

\begin{document}

\begin{tcolorbox}[colback=yellow!10!white, colframe=yellow!50!black, boxrule=0.5pt, width=\textwidth]
\centering \textbf{This is a preprint version of the article accepted for publication in the Journal of Systems and Software. The final published version will be available via Elsevier.}
\end{tcolorbox}

\vspace{1em}

\begin{center}
\includegraphics[width=\textwidth]{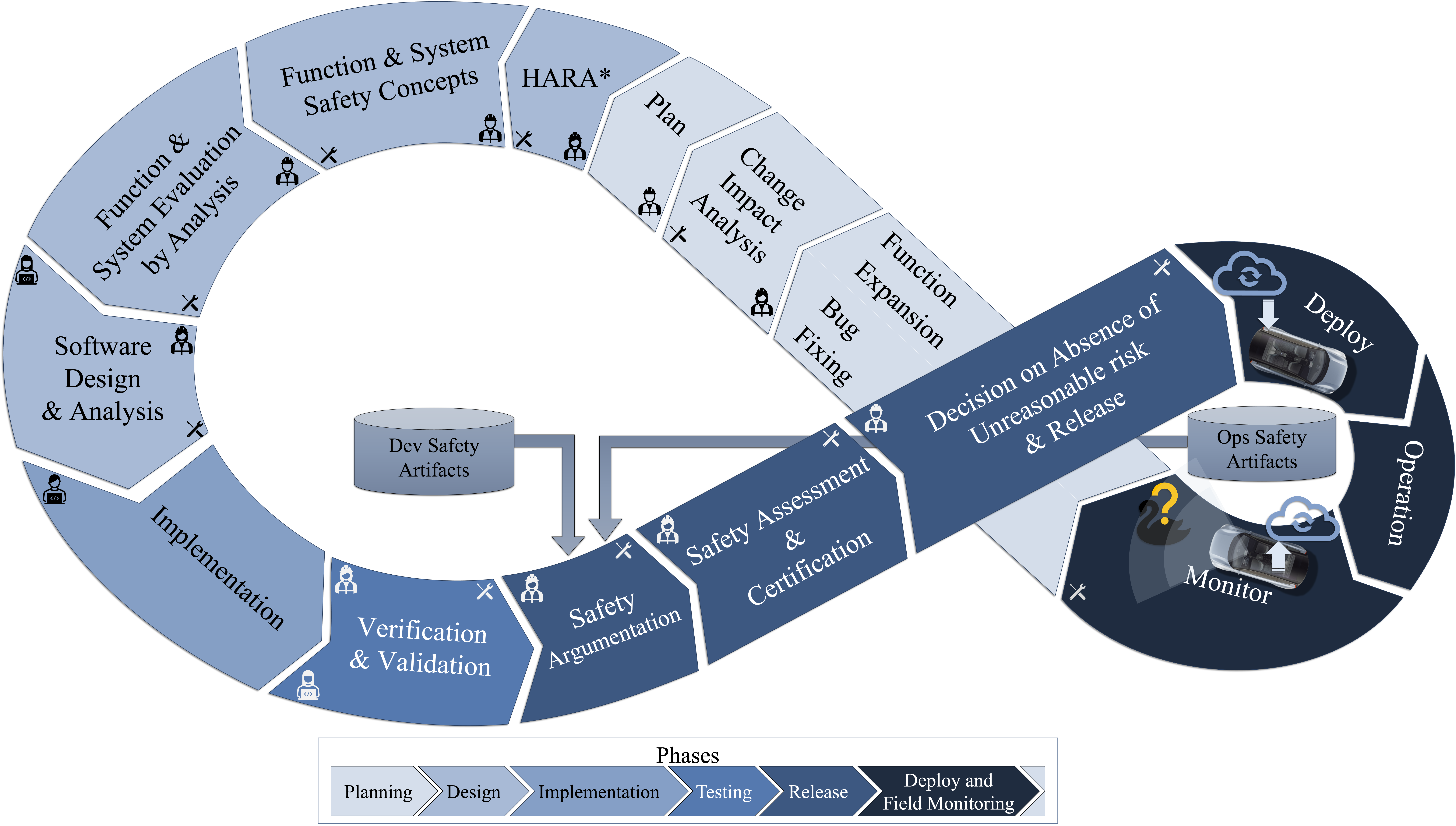}

\end{center}

\title{The DevSafeOps Dilemma: A Systematic Literature Review on Rapidity in Safe Autonomous Driving Development and Operation}

\author[label1,label3]{Ali Nouri} 
\author[label1,label2]{Beatriz Cabrero-Daniel} 
\author[label3]{Fredrik Törner} 
\author[label1,label2]{Christian Berger} 

           
\affiliation[label1]{{Chalmers University of Technology, Department of Computer Science},
            city={Gothenburg},
            country={Sweden}}

\affiliation[label2]{{University of Gothenburg, Department of Computer Science},
            city={Gothenburg},
            country={Sweden}}

\affiliation[label3]{{Volvo Cars},
            city={Gothenburg},
            country={Sweden}}
\begin{abstract}
Developing autonomous driving (AD) systems is challenging due to the complexity of the systems and the need to assure their safe and reliable operation. 
The widely adopted approach of DevOps seems promising to support the continuous technological progress in AI and the demand for fast reaction to incidents, which necessitate continuous development, deployment, and monitoring.
We present a systematic literature review meant to identify, analyse, and synthesise a broad range of existing literature related to usage of DevOps in autonomous driving development.
Our results provide a structured overview of challenges and solutions, arising from applying DevOps to safety-related AI-enabled functions. 
Our results indicate that there are still several open topics to be addressed to enable safe DevOps for the development of safe AD.
\end{abstract}

\begin{frontmatter}

\begin{highlights}
\item Applying DevOps to autonomous driving has presented several open topics to be addressed.
\item DevSafeOps is introduced, incorporating safety-related activities into DevOps iterative loops.
\item Our systematic literature review identifies 11 clusters of challenges for essential safety activities in the DevSafeOps loop.
\item Potential solutions are identified from the literature and mapped to challenges in DevSafeOps.

\end{highlights}


\end{frontmatter}

\renewcommand\arraystretch{1.3}

\section{Introduction}

Artificial Intelligence (AI) has seen widespread adoption in various fields over the past decade, including the automotive industry. Autonomous Driving (AD) and Advanced Driver Assistance Systems (ADAS) are currently two prominent scenarios to apply AI. Ensuring the robustness and safety of such AI-based systems in a dynamic environment is a complex and multifaceted process that requires continuous monitoring and fast yet systematic reactions to the possible identified hazards.

AD development started back in the 1980s~\cite{gudla2022review} and there have been local experiments; however recent loss events highlight its ongoing safety challenges.
For instance, in a recent mishap involving a robotaxi, a pedestrian was severely injured~\cite{cruise2024, koopman2024anatomy}.
According to the investigation report~\cite{cruise2024}, the cause was neither hardware nor software failure. The AD perception detected both the pedestrian and the adjacent vehicle.
Weak recognition and response to nearby incidents, along with an inaccurate post-crash world model, are some of the technical issues and challenges highlighted by Koopman~\cite{koopman2024anatomy}. Moreover, a human safety driver is also considered a potential solution for handling these unforeseen complex scenarios~\cite{koopman2024anatomy}. This incident highlight the need for novel approaches in the design of ADS that enable the system to adapt its behavior to similar unforeseen, complex, and yet numerous events. Releasing the system without sufficient confidence and necessary fallback strategies not only leads to safety risks but also poses delays in deployment or termination of the project.
For instance, the aforementioned example led to the immediate revocation of ADS deployment and testing permits by the California Department of Motor Vehicles (DMV).~\footnote{DMV STATEMENT ON CRUISE LLC SUSPENSION, accessed February 28, 2025, \url{https://www.dmv.ca.gov/portal/news-and-media/dmv-statement-on-cruise-llc-suspension/}} 

The innovation rapidity in the automotive industry necessitates companies to adopt continuous development and integration approaches to remain competitive.
Continuous approaches, such as DevOps, aim at function growth and refinement to lead to better customer experience after each design iteration~\cite{devopsgoogle}. 
Hence, DevOps has the potential to enable continuous loops of monitoring and software adaptation for AD. It facilitates the expansion of the Operational Design Domain (ODD). Additionally, DevOps is crucial for maintaining the safety of the system against detected anomalies, and improve the adaptation speed to context/environmental/regulation updates~\cite{zeller2024toward, weiss2024approach}.
Hence, the safety community aims to integrate some aspects of DevOps into standards such as ISO/PAS 8800 (Safety and Artificial Intelligence) and ISO/TS 5083 (Safety for Automated Driving Systems).

However, the safety process required by automotive standards such as ISO 26262~\cite{ISO26262} or ISO 21448~\cite{sotif}, and regulations such as UNECE R157 (ALKS)~\cite{ALKS} can make it challenging to utilise full capacity of DevOps.
 These standards prescribe engineering rigour for increasing integrity and reducing safety risks, which are essential in safety-related domains. On the other hand, the fundamental concept of DevOps is to emphasize speed and rapid updates. This can be perceived as a dilemma as speed and safety seem to not be able to be combined at a glance.
 As accident reports show, this dilemma has, in some cases, been resolved in favor of speed and innovation rather than safety. One notable example occurred in the avionics industry, where prioritizing speed and competitiveness over thorough safety processes during the modification of existing aircraft models not only led to fatalities but also resulted in the worldwide grounding of that model~\cite{defazio2020final}.
 
 However, DevOps does not dictate an absolute level of speed; rather, it is a mindset focused on continuous improvement and adaptability in relative terms, compared to earlier practices within an organization or even external organizations. Hence, DevOps applied in safety related domains implies that the integrity and safety is viewed as prerequisites, but within these boundaries the DevOp principles are applied to its full extent.

To the best of the author's knowledge, there is no systematic attempt to synthesise the research on this challenge targeting in particular the automotive domain. This study aims at reviewing the literature on safety-related DevOps challenges, which are considered to be applicable for AD, and explores proposed solutions and mitigation strategies by answering the following research questions:

\begin{itemize}
    \item \textbf{RQ1.} What challenges are identified in literature when applying DevOps to safety-related AD functions? 
    \item \textbf{RQ2.} What solutions are proposed in literature for these challenges while fostering rapidity and safety? 
    \item \textbf{RQ3.} What challenges for DevOps in safety-related AD applications still remain open in literature? 
\end{itemize}

We followed a systematic approach, based on the work by Kitchenham and Charters~\cite{Kitchenham}, to synthesise a broad range of literature on DevOps engineering process in safe AD development. The results map current challenges of DevOps for safety-related AI-enabled functions (RQ1) to suggested state-of-the-art solutions (RQ2). The analysis also reveals several opportunities for research to better address the identified challenges (RQ3).

The rest of this paper is organised as follows.
Background and related work on DevOps and safety-related AI-enabled functions are presented in Section~\ref{sec:background} and Section~\ref{sec:RelatedWork}. The research design and implementation are described in Section~\ref{sec:methodology}. The results of the study are presented and discussed in Section~\ref{sec:results} and Section~\ref{sec:discussion} concludes the paper.

\section{Background}
\label{sec:background}

\subsection{Safety - Definitions, Regulations and standards landscape}
To assure and argue the safety of ADS, the `safety criteria' shall first be defined, both qualitatively and quantitatively.
Several stakeholders are involved in defining the criteria, including Consumers, industries involved in the ADS development, and legislators.

United Nations Economic Commission for Europe (UNECE) published UN Regulation No. 157~\cite{ALKS}, which defines qualitative and quantitative criterion to ensure the safety of Automated Lane Keeping Systems (ALKS). As a qualitative requirements, UNECE R157 recommends compliance with ISO 26262~\cite{ISO26262}, ISO 21448~\cite{sotif}, and ISO 21434~\cite{isocs} by requiring a competent auditor and assessor in these standards.

ISO/TR 4804~\footnote{Road vehicles - Safety and cyber-security for automated driving systems - Design, verification and validation}~\cite{isotr4804} defines it as the Absence of Unreasonable Risk (AUR). 
ISO 26262, ISO 21448, and ISO 8800 refine the definition (i.e., AUR), making it more specific based on the root causes they target:

\begin{enumerate} [leftmargin=2.2cm]
 \item[ISO 26262:] ``AUR due to hazards caused by \textbf{malfunctioning behaviour of E/E systems}''
 \item[ISO 21448:] ``AUR due to hazards resulting from \textbf{functional insufficiencies of the intended functionality} or \textbf{its implementation}.''
 \item[ISO 8800:] ``AUR due to \textbf{errors of the AI system}.''
\end{enumerate}

Positive Risk Balance (PRB), which requires ADSs to ``cause fewer crashes on average compared to those made by drivers,'' can be employed as a quantitative safety criterion~\cite{isotr4804} and acceptance criterion in ISO 21448~\cite{sotif}. However, this raises questions about ``how many less?'' and ``which driver?''~\cite{koopman2023breaking}. As highlighted by Koopman and Widen, both AUR and PRB concepts fall short for ADS, since they fail to account for scenarios requiring moral and contextual understanding and might lead to risk redistribution to other road users~\cite{koopman2024redefining}.

ISO 26262~\cite{ISO26262} (functional safety, or so-called FuSa) provide process requirements to avoid or mitigate systematic failures in hardware and software, as well as random hardware faults, during the design, implementation, verification, validation, and field monitoring phase. The resulting work-products from these activities are then used as evidences in a safety case to argue the achievement of functional safety for the item.

As a complementary standard to ISO 26262, ISO 21448 (Safety of Intended Functionality, or SOTIF)~\cite{sotif} contains requirements and recommendations for avoiding hazardous events caused by functional insufficiencies, incorrect or inadequate Human-Machine Interface (HMI), and insufficiencies in Artificial Intelligence (AI)-based algorithms.
SOTIF requires the definition of acceptance criteria, which can be both qualitative and quantitative (e.g., one event per X km), as the first level of safety requirements. The fulfillment of each acceptance criterion is then argued based on evidence from analysis and testing for each validation target (e.g., no hazardous behavior during X hours of testing)~\cite{sotif}. If any of the acceptance criteria are not met, a functional modification (i.e., changes in design and specification) is needed.

ISO 4804~\cite{isotr4804} and the upcoming ISO 5083 are ADS-specific standards that discuss additional process aspects of ADS in relation to ISO 26262 and ISO 21448. Additionally, they provide technical and architectural recommendations for ADS.
UL 4600 is another ADS-specific standard that focuses on system-level aspects~\cite{KoopmanUL4600}. There are also general safety standards, such as IEC 61508~\cite{IEC61508}, which serve as the foundation for other domain-specific safety standards such as ISO 26262 and ISO 21448.

Employing AI technology in automotive software is not specific to ADS. However, relying on it without human supervision for safety-related tasks in a complex environment presents a new challenge for the automotive industry, which ISO 8800~\footnote{ISO/CD PAS 8800, accessed May 20, 2024, \url{https://www.iso.org/standard/83303.html}} aims to address.

National Highway Traffic Safety Administration (NHTSA)~\cite{NHTSA2017VisionForSafety} encourages companies to submit a publicly available Voluntary Safety Self-Assessment (VSSA) report, which outlines how they address safety concerns, including system safety, ODD, validation methods, and data recording. While this is not mandatory, the VSSA report aims to promote transparency and build public trust. As of today, 28 companies have a link to their VSSA or their safety page describing their key safety design elements, which are analyzed in our study from an industrial perspective.

\subsection{Abstraction Levels}
\label{subchap:AbsLevl}
Automotive system engineering requires a well-established modular and multi-abstraction architecture and requirements engineering process. Each abstraction level consists of multiple modules, each of which is abstracted by removing or merging unnecessary characteristics of the elements in the module~\cite{SAESTPA}. Traceability between the elements and requirements at the most abstract level and the most granular level is essential to enable impact analysis and maintainability of the system after modifications during the life cycle of the system.
The recommended abstraction levels are as follows, but are not limited to these, as they depend on the scope and complexity of the product:

\textbf{Concept} phase starts with a short description of the item, describing its functionality and ODD, known as the `Item Definition'~\cite{ISO26262}. 
ODD is a set of predefined restrictions on the function due to its limitations to ensure safety~\cite{stettinger2024trustworthiness}.
It also includes the item's boundary diagram, which represents its interaction with the environment, operator, and other agents.
`Hazard Analysis \& Risk Assessment' (so called HARA in the context of ISO 26262~\cite{ISO26262}) and Hazard Identification \& Evaluation (in the context of ISO 21448~\cite{sotif}) are then employed to identify all hazardous events that result from the occurrence of malfunctions or functional insufficiencies in a specific scenario. A safety goal (ISO 26262~\cite{ISO26262}) or an acceptance criterion (ISO 21448~\cite{sotif}) is specified for the safety-related hazardous events to avoid or mitigate.

\textbf{Function} level is the next abstraction level, which includes the functional architecture, inductive or deductive safety analysis methods and functional safety requirements all gathered in a functional safety concept~\cite{ISO26262} and a SOTIF safety concept~\cite{sotif}. Ideally, the functional/SOTIF safety requirements are technology-agnostic.

\textbf{System} level is the first level where technical aspects of a function are designed. A technical safety concept is the container for system-level (technical) safety requirements resulting from design choices based on the safety analysis results.

\textbf{Software} and hardware levels are the last abstraction levels.
Software safety requirements are derived from system level technical safety requirements employing safety analysis and the safety solutions are designed in the software architectural design. Depending on the complexity of the system, these levels can be further broken down into more levels due to the complexity of the product or other constraints.

\section{Related Work} \label{sec:RelatedWork}

A limited body of academic literature addresses some of the challenges of integrating DevOps practices in the development of safety-related software. Moreover, there is a growing set of literature that recognises the difficulties of integrating AI into such systems. These challenges are illuminated by grey literature, multi-vocal literature reviews, or interviews such as ~\cite{Nouri2022Experience}.

DevOps couples Continuous Development (Dev) and Operations (Ops)~\cite{24safeopsbosch}. Primary needs include monitoring, diagnostics, and event recording, satisfied by triggers and data loggers~\cite{24safeopsbosch,25fayollas2020safeops}. 
Using DevOps in an automotive engineering context can enable rapid communication towards original equipment manufacturers (OEMs) to allow new software releases, enabled by over-the-air updates~\cite{25fayollas2020safeops, czarnecki2019software,kugele2022architecture, Nouri2022Experience}, which may be even more rapid by 5G and 6G~\cite{kugele2022architecture}. Flexible system architectures such as central computing units allow for safe integration of new software without compromising the safety of current mature software~\cite{25fayollas2020safeops}. The UP2DATE~\cite{Agirre2020Agile} European project aimed to propose new software update architectures for mixed-criticality systems.

Machine Learning (ML) is a crucial technology in AD, especially for supporting perception tasks. However, unbounded complexity of the environment and distributional shift~\cite{borg2018safely} requires continuous monitoring, data gathering and improvement during operation (i.e., MLOps)~\cite{18siddique2020safetyops, 32simonCIA, sotif, MLOpsgoogle, john2023towards}. Hence, successful ML development requires an iterative, close, and rapid connection between operation and development~\cite{borg2021aiq}.

The SafeOps approach by Munk and Schweizer focuses on ensuring safety in iterative development processes~\cite{24safeopsbosch}. Similarly, the Safe\underline{ty}Ops framework by Siddique addresses challenges of integrating safety and DevOps and emphasises the need to adapt DevOps practices to meet the unique needs for safety~\cite{18siddique2020safetyops}.
Considering the provided terms and definitions, we propose DevSafeOps to refer to the integration of safety and DevOps.
The absence of ``Dev'' in the other two terms (i.e., SafeOps~\cite{24safeopsbosch, 25fayollas2020safeops}, SafetyOps~\cite{18siddique2020safetyops}) make them less suitable compared to ours, where the concept of ``safe by design''~\cite{galbas2024vv} plays an important rule. Moreover, the term SafeOps is used in other contexts, such as safe military operations~\cite{kott2011safe}.

To the best of the authors’ knowledge, none of the aforementioned studies systematically focuses on and reviews DevOps challenges for AI-enabled safety-related automotive functions such as AD with the goal of mapping them to state-of-the-art solution suggestions. 
However, the literature does report useful knowledge for the field on specific topics like verification and validation (V\&V)~\cite{borg2018safely}, agile methodologies in safety~\cite{myklebusta16agile}, or cyber-security~\cite{sanchez2020security}. 
Thus, the present paper aims to fill this gap, providing a comprehensive view on rapidity in safe AD DevOps, as described below.

\section{Method: a Systematic Literature Review} \label{sec:methodology}

The present study uses a Systematic Literature Review (SLR) in order to explore the current landscape of rapidity for developing safe AD DevOps. Prior to the start of the study, we queried Google Scholar to retrieve papers containing the keywords ``DevOps,'' ``Safety,'' and ``Systematic Literature Review.'' This pilot search demonstrated that there is no available SLR on safety of DevOps, since the retrieved papers focus instead on other aspects such as quality~\cite{cespedes2020software} or cyber-security~\cite{sanchez2020security}.

After this preliminary study, we designed a methodology to gather, analyse and synthesise relevant literature on the topic at hand as described below. Given the high industrial relevance of this endeavour, we opted to include grey literature and non-peer reviewed articles, which enriches the results.

\subsection{Data Sources and Search Strategy}

\begin{figure*}
    \centering
    \includegraphics[width=\linewidth]{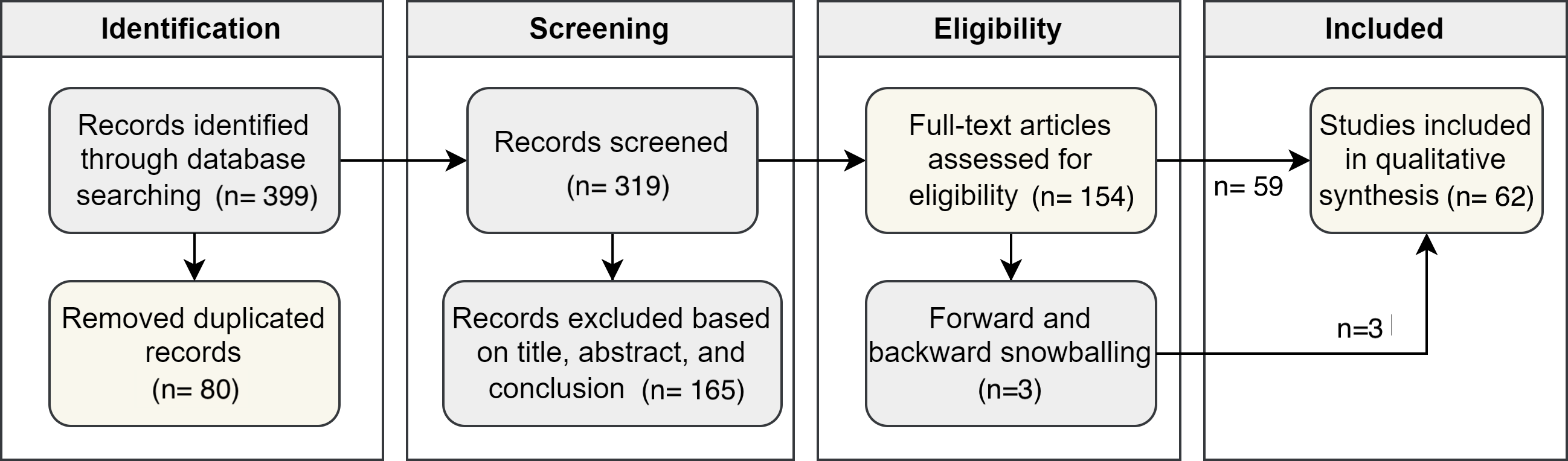}
    \caption{Flow diagram depicting the application of inclusion and exclusion criteria.}
    \label{fig:flow}
\end{figure*}

To begin this process, we conducted preliminary searches to identify existing systematic reviews and to assess the volume of potentially relevant studies. With this analysis, we identified which keywords are used within each relevant research topic, and what are boundaries of the existing literature. This information was used to design a query to retrieve existing literature on safety and rapid software development. 
For the selection of primary studies, we queried four literature databases (ACM Digital Library, IEEE Xplore, and Scopus) and Google Scholar, that indexes also pre-print servers such as arXiv, using the following query: 

\begin{center}
\noindent\texttt{safety AND (26262 OR 21448 OR SOTIF)\\
AND (DevOps OR MLOps) AND\\
(((autonomous OR automated OR driverless) AND
(vehicle OR drive OR driving OR car)) OR self-driving)}
\end{center}

The result of this query was then assessed by the first and second authors to identify the primary studies to be evaluated and synthesised in this study, as reported in Figure~\ref{fig:flow}.

\subsection{Study Selection}

Selection criteria were defined to dictate whether a study would be included or excluded from the set of primary studies to analyse. The design of these criteria was based on preliminary exploratory efforts. For instance, given that older studies seemed to discuss fundamental yet still relevant challenges, no study was excluded because of its publication date. As a result, the earliest identified publication is from 2015. As a result, retrieved papers are included if they meet the following inclusion criteria:

\begin{enumerate} [leftmargin=0.95cm]
    \item[\textbf{Inc.1}] Written in English.
    \item[\textbf{Inc.2}] Presenting challenges, problems, and confounding factors in applying DevOps (or MLOps) to safety-related functions in automotive, or discussing practices, guidelines, lessons learned, methods, algorithms, approaches, or tools to address the identified challenges.
\end{enumerate}

\noindent The ``exclusion criteria'' are defiend as following:

\begin{enumerate} [leftmargin=1cm]
    \item[\textbf{Exc.1}] Does not discuss about safety of DevOps or MLOps in the automotive domain.
    \item[\textbf{Exc.2}] Does not report on any challenges or solutions.
    \item[\textbf{Exc.3}] Is not written in English.
\end{enumerate}

The first author acted as the main assessor evaluating each prospective primary study, while the second author independently evaluated all included records, and explored a sample of the excluded studies (random selection and most cited ones) to validate the exclusion rational. There were no discrepancies between the evaluations made by the first and second authors. Figure~\ref{fig:flow} summarises the selection process at each stage.

\subsection{Data extraction and synthesis}

After removing duplicate records from the results of the database search, we screened the records based on title, abstract, and conclusion, using the aforementioned criteria (screening phase). Then, in the eligibility phase, the full text was analysed and relevant records were selected. If additional records were identified through snowballing, they were also included (see Supplementary Materials\footnote{Supplementary Material regarding included and excluded data, and the rationale, \url{https://zenodo.org/records/14946487}}).

The goal was to extract information related to applying DevOps in safety-related aspects of AD or applicable to AD. Retrieved information includes the first author, publication year, publication venue, research type, and industrial sector. Moreover, data to address each research question was retrieved, i.e., challenges identified (RQ1), proposed software technologies, and recommendations to address specific issues (RQ2), and open challenges and future work directions (RQ3).

Then, thematic analysis was performed on the primary studies. First, papers were read to form initial ideas for analysis and generating inductive codes from lists of challenges, practices, and factors. These codes were combined into potential themes, later refined by merging, balancing, or dropping codes. By doing so, we grouped the challenges and solutions based on safety activities and relevant categories. Clear and concise terms for each challenge and solution were defined to present the results in Section~\ref{sec:results}.

In addition, we aimed at aggregating the extracted information to derive clusters of
challenges and solutions, our work also gets a notion of a mapping study. 
In that regard, we aimed at maintaining the integrity of the original information by only linking challenges to solutions if they are mapped in the original authors' publication. We acknowledge, though, that some solutions could be applied to multiple challenges and some challenges may overlap with other groups. For that reason, we clearly indicate and justify solutions that are not mapped to a specific challenge in literature. Furthermore, we provide a mapping aimed to address \textbf{RQ3} that combines existing studies and serves as a basis to discuss the interrelation and inter-dependency of the reported challenges.

\subsection{Complimentary Studies: the Industrial Perspective}

In order to highlight the importance of industrial contributions to this field of study, we mark any cited material involving industrial authors with [X]$^{I}$.
Moreover, a complementary set of reports, produced by industrial parties, was analysed and included alongside the retrieved papers, as ``Industrial Context'': We analysed all Voluntary Safety Self-Assessment (VSSA) reports from companies involved in ADS development, which are published on the NHTSA website~\footnote{NHTSA Voluntary Safety Self-Assessment, accessed February 28, 2025, \url{https://www.nhtsa.gov/automated-driving-systems/voluntary-safety-self-assessment}}. Snowballing was also performed from these reports to identify other relevant reports or publications.

\subsection{Threats to validity}

Different types of information sources have been used, due to the nature of the research in the scope of the automotive industry.
Therefore, the clustering could be sensitive to the way that the review has been conducted.
For instance, safety case, assessment, certification, and impact analysis challenges are related to more granular safety activities due to their interconnection in different abstraction levels, which could lead to other possible clustering patterns.
To mitigate this, we followed the structuring pattern used in ISO 26262, which forms the foundation of automotive safety. Moreover, this clustering was done for presentation and communication purposes, and it does not influence the data extraction process or the conclusions drawn. Lastly, since the clustering was applied only after data synthesis, it does not introduce bias or affect the internal validity of the study.

While solutions from other domains can be used in automotive, we focused on prescribed approaches for automotive, as otherwise, such potential approaches require to be validated if applicable to such standards, which is outside the scope of this literature review. Moreover, the automotive domain has its own constraints and characteristics such as complexity, competitiveness, speed, and distributed development way of working, which might be different in other industries.

This work might be missing or excluding some relevant studies on safety-related challenges in continuous development approaches. To mitigate this, we used different digital libraries with a query that was designed iteratively based on the independent initial findings of the authors with the goal to complement each other. Moreover, snowballing was used (see Figure~\ref{fig:flow}) to screen other related studies, strictly following the pre-defined review protocol in Section~\ref{sec:methodology}, defining the inclusion and exclusion rationales. Moreover, additional sources, mainly grey literature and AD developer's VSSA reports, have been added to cover potentially existing gaps in academic literature. 
The second author performed a cross-check on all the included papers and a randomly selected set of the excluded studies (25\% over the final set, corresponding to one every four of the papers listed in the Supplementary Materials) to reassess whether they had to be included or not in each stage of the screening process.

\section{Results} \label{sec:results}

\begin{sidewaysfigure}
  \includegraphics[width= 1\textwidth]{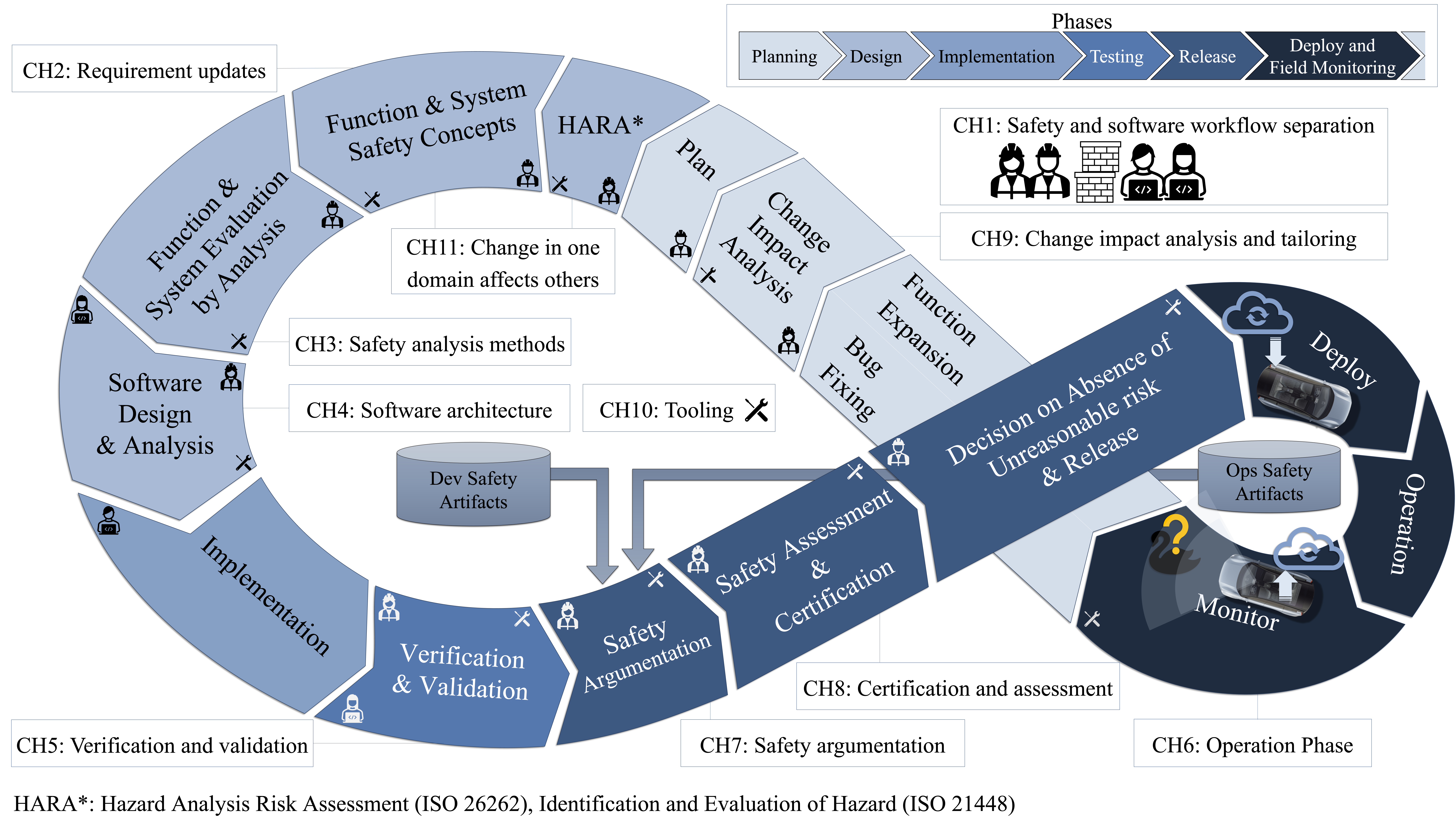}
  \caption{The figure outlines the safety activities that are necessary to be carried out in DevSafeOps cycles, along with the corresponding challenges identified in this study. The asymmetrical shape of the loop illustrates the significant number of activities required on the development side, as compared to the operation phase. The loop starts from planning in the first iteration, and in the successive iterations, it begins with Function expansion or bug fixing. The artefacts created during the Development and Operation phases of DevSafeOps are continuously collected to form the safety argumentation. The intensity of the colour represents different phases in the DevSafeOps loop.}
  \label{fig:DevOps}
\end{sidewaysfigure}

This section describes the synthesis of reviewed papers, aimed to answer our research questions. To this end, challenges ($CH_n$) and proposed solutions ($Sol_{n,m}$) are listed here, grouped based on the results of our thematic analysis.
We integrated the required safety activities into our interpretation of the DevOps infinity loop, which is shown in Figure~\ref{fig:DevOps}, and then mapped all identified challenges to the activities.
To structure the findings from the reviewed papers, they were clustered into 11 challenges, each aligned with key activities or phases defined in ISO 26262.
 For each challenge we first present our own analysis regarding the relevance of proposed solutions and other possible approaches are discussed at the end of each challenge in ``\emph{Discussion}''. Finally, for each challenge, the findings and analysis results from VSSA reports are presented in ``\emph{Industrial Context}''.
 Our concluding interpretations and reflections are discussed in Section~\ref{sec:discussion}.

\vspace{0.3cm}
\begin{tabular}{|p{0.9\columnwidth}|}
  \hline
  \textbf{CH1: Safety and software workflow separation} \\  
  \hline
\end{tabular}
\vspace{0.1cm}

According to Siddique in his 2020 position paper ``SafetyOps'', safety activities are traditionally expected to be performed independently from design activities~\cite{18siddique2020safetyops}$^{I}$. According to Fayollas et al., such practice may also result in conflicts between the development team and safety experts~\cite{25fayollas2020safeops}$^{I}$. As a consequence, safety activities have been primarily managed by safety experts, with little engagement from the development team. This is specifically true for Agile teams~\cite{20warg2019continuous}$^{I}$, which results in misalignment and delays in releases. The safety experts referred to here are those engaged in producing safety artifacts and designs, while confirmation review tasks (e.g., assessment) necessitate separation and independence.

\emph{Sol1.1:} Hanssen et al. proposed ``SafeScrum'' for Agile Development
of safety-related Software~\cite{hanssen2018safescrum}$^{I}$, introducing additional roles with respect to traditional Scrum that are required to meet the requirements of IEC 61508~\cite{IEC61508}. A ``Quality Assurer'' is a Scrum team member with relevant safety knowledge who ensures that software quality assurance tasks, the safety plan, and the safety validation plan are completed~\cite{hanssen2018safescrum}$^{I}$, ~\cite{myklebust2020agile}$^{I}$.

\emph{Discussion:} Potential root cause of this challenge is the lack of safety knowledge within the development team in some companies. Following the SafeScrum~\cite{hanssen2018safescrum}$^{I}$ proposal can help to address this issue by including at least one safety expert in the team, ensuring the quality of safety activities, and guiding the team in performing the tasks. The daily stand-up meetings facilitate collaboration between developers and safety experts, enabling them to update each other and correct misalignment faster. This, together with knowledge sharing (T-shaped skills) and training sessions, could increase the safety skills of the team. 
Independent tester, ``reliability, availability, maintainability and safety (RAMS)'' engineer, and independent safety assessor are external to the team, with I2 and I3 degrees of independence~\cite{hanssen2018safescrum}$^{I}$. 
The independent safety assessor, who is not part of the product development or the same organization (i.e., I3 degree of independence), receives evidence of compliance and ensures that the requirements from the standard are fulfilled.

\emph{Industrial Context:} According to IKE's Safety Report~\cite{IKE-VSSA}, it is the responsibility of the systems engineering team to ensure the safety of their work and the potential implications of their engineering decisions. Their responsibilities also include peer-reviewing software before commits and performing safety analyses, such as STPA.

\vspace{0.3cm}
\begin{tabular}{|p{0.9\columnwidth}|}
  \hline
  \textbf{CH2: Requirement updates} \\  
  \hline
\end{tabular}
\vspace{0.1cm}
 
Requirements engineering for safety-related applications typically consists of multiple abstraction levels, starting with safety goals in the context of ISO 26262 or validation targets and acceptance criteria in ISO 21448 and then, are broken down into more granular ones~\cite{Nouri2022Experience}$^{I}$. 
The DevOps approach starts with a Minimum Viable Product (MVP) that is iteratively improved by adding features~\cite{14KrystofREAD}, and then new features are added to the product to improve the user experience continuously.
Then it is necessary to update/remove legacy requirements and to add new ones in each iteration. This made traditional requirements engineering inadequate in capturing benefits of rapid DevOps cycles~\cite{1jantango}. Bosch et al. report that requirements are typically agreed before the implementation phase, and sudden changes are avoided because they are estimated to be costly~\cite{1jantango}. Incorporating multiple confirmation measures such as verification reviews, confirmation reviews, or assessments in requirements engineering result in even more needed effort and potential delays.
According to Habibullah et al., the inclusion of ML elements makes it challenging to employ requirement traceability to trace the source of issues. Therefore, it is necessary to consider requirement traceability for models and datasets as well~\cite{habibullah2024requirements}$^{I}$. For instance, in AD, it is also important to link safety goals to scenarios.

\emph{Sol2.1:} AI- and data-driven development are proposed as two potential alternatives to traditional requirements engineering~\cite{1jantango}, ~\cite{habibullah2024requirements}$^{I}$. \textit{AI-driven development} is widely used in development of perception software of AD, such as object detection, by using ML techniques. ML is trained on large data-sets of pictures or point clouds of objects instead of relying on detailed requirements about shape and appearance of an object. \textit{Data-driven development} enables the team to use data in addition to requirements to develop software, and then use metrics to evaluate the performance of the developed software~\cite{1jantango,kugele2022architecture}, ~\cite{hartmannsgruber2024improving}$^{I}$. In both cases, confidence in system performance is initially low due to insufficient data.
Thus, by introducing limitations and a conservative approach, safety is maintained. Later, supported by data, unnecessary constraints can be identified and removed~\cite{warg2024salience4cav}$^{I}$, ~\cite{20warg2019continuous}$^{I}$. Continuous data collection and monitoring of the system and the environment through event detectors and monitoring tools help to identify emerging edge cases~\cite{14KrystofREAD}. 

\emph{Discussion:} The use of data as an input for ML-based software development is discussed in ISO 21448~\cite{sotif} and upcoming standards such as ISO/TS 5083 and ISO/PAS 8800. Nevertheless, it is not yet explored as a replacement for safety-related requirements at any abstraction level. 
Moreover, although ML-based software is utilised in perception systems, where data is employed as a requirement for software behaviour, there are still requirements on the utilised data and ML technology. These requirements are in turn used in the safety V\&V process of the entire perception system and its components. 

\emph{Industrial Context:} According to VSSA of Torc~\cite{Torc-Robotics-VSSA}, ODD evolution over time is decomposed into testing ODD (T-ODD) and ODD expansions (eODDs). T-ODD refers to the ODD that the current software is capable of handling and is being tested for, while eODD represents the ODD planned for the next phase of software development and expansion.
Finally, after multiple evolutions, the deployment ODD is formed, indicating the ODD of the deployed software.
However, the report does not describe the impact of this iterative ODD expansion on the evolution of safety requirements or other artifacts.

\vspace{0.3cm}
\begin{tabular}{|p{0.9\columnwidth}|}
  \hline
  \textbf{CH3: Safety analysis methods} \\  
  \hline
\end{tabular}
\vspace{0.1cm}

``Deductive safety analysis'' (Top-Down) methods are employed to identify the potential causes of the higher level safety requirement violation.
 For instance, Fault Tree Analysis (FTA) can be employed to identify the potential root causes for requirements in higher abstraction level (e.g., safety goals violation) in the current abstraction level (e.g., functional modules).
 
``Inductive safety analysis'' (Bottom-Up) methods are used to identify the effects of a specific failure mode or performance limitation on higher abstraction levels. If the effect is related to a violation of safety requirements at higher levels, then the failure mode or performance limitation shall be avoided or mitigated. For instance, Failure Mode and Effect Analysis (FMEA) approach starts from the failure modes of each module at the current level (e.g., function level) and identifies the effects at higher levels (e.g., violating safety goal).

Analysis methods such as ``System Theoretic Process Analysis'' (STPA), ``Fault Tree Analysis'' (FTA), and ``Failure Mode and Effect Analysis'' (FMEA) are performed manually and separately without considering other methods~\cite{18siddique2020safetyops}$^{I}$ ~\cite{AliSEAASTPA}. This can lead to inconsistency between the results of different analysis methods. Moreover, safety analysis shall be updated whenever a change occurs in the system or new evidence and events are collected from the field; otherwise, the results may become outdated or incorrect~\cite{thomas2023toward}.
Reviewing and updating of them is a manual and time consuming task. This may not only cause delays in releasing new software or system~\cite{18siddique2020safetyops}$^{I}$, but also raises the risk of late reactions to field data, which might lead to an accident.
Additionally, the elements in each method (e.g., failure mode, architectural elements, effects) should be traceable across different abstraction levels, enabling the tracing of causes to the most granular level~\cite{AliSEAASTPA}$^{I}$.

\emph{Sol3.1:} Integrating all safety analysis methods into a single framework and connecting the data to the relevant parameters enable automatic updating of said parameters with each data update~\cite{18siddique2020safetyops}$^{I}$. For instance, in this way the frequency of failures in FMEA can be linked directly to field data~\cite{18siddique2020safetyops}$^{I}$ and being updated automatically instead of traditional manual failure calculation using an outdated failure catalogue.
This might require adaptations of methods, such as the proposed tailoring of STPA in~\cite{AliSEAASTPA}$^{I}$, to enable modularisation and multi-abstraction levels while maintaining traceability between elements down to the most granular level.
Using unified causal model (i.e., cause and effect relationships) for safety analysis can be a potential solution. Maier et al. proposed a framework for integrating causal models into DevOps, called CausalOps~\cite{maier2024causalops}.
Similarly, Hybrid Causal Logic (HCL) methodology~\cite{thomas2023toward} proposes a data-driven and dynamic safety analysis. In this approach, safety analysis is integrated with Event Sequence Diagrams (ESD) and Bayesian Networks, enabling continuous and dynamic improvements. HCL leverages Bayesian Networks to update safety analysis results based on new field data. Moreover, ESD integration ensures that all possible failure paths and scenarios are considered in the analysis.

\emph{Sol3.2:} Zeller et al.~\cite{zeller2024toward}$^{I}$ posed the question of whether it is possible to automate hazard identification and risk assessment, a challenge that has been addressed and implemented by~\cite{nouri2024RELLM}$^{I}$.
As safety requirements are in natural language, hence Natural Language Processing (NLP) can be a potential solution, an idea that has been technically implemented. 
The impressive performance of Large Language Models (LLMs) in natural language understanding and text generation has led to their use in requirements engineering and safety analysis~\cite{nouri2024RELLM}$^{I}$ ~\cite{Nouri2024CAINLLM}$^{I}$. Nouri et al.~\cite{nouri2024RELLM}$^{I}$ designed a pipeline consisting of LLM agents (GPT-4 with specific prompt templates), each specialized in a subtask of HARA, communicating with one another to deliver a HARA for any function, including safety goals.
These models can also be used for data collection and preparation in natural language from various sources, such as police accident data containing challenging scenarios specific to each country.
For instance, B{\"a}umler et al. used BERT to classify German police accident data into 30 different turning accident types~\cite{baumler2024predicting}.

\emph{Discussion:} Data-driven development, discussed in \emph{Sol2.1}, is a prerequisite for \emph{Sol3.1}. This emphasises the importance of integrating data in every aspect of the development process such as requirements engineering and analysis activities.
As mentioned in \emph{Sol3.2}, the use of NLP techniques is not limited to safety analysis and can be expanded to other activities documented in natural language. However, the results of these studies must be validated rigorously with special attention. For instance, as highlighted by~\cite{nouri2024RELLM}$^{I}$, testing LLMs against existing open-source baselines only assesses their performance on data they might have been trained on and does not validate their performance for new input.

\emph{Industrial Context:} Safety analysis is a ``living'' activity throughout the development cycle and should be updated based on field data~\cite{IKE-VSSA}. For instance, near-loss incidents should be analysed to update the system or software and to prevent future failures~\cite{IKE-VSSA}. 
Moreover, in analysis methods such as STPA, failure modes related to the software update process (e.g., incomplete updates or incorrect versions) should also be analysed, and corresponding mitigation strategies can be developed~\cite{IKE-VSSA}.

\vspace{0.3cm}
\begin{tabular}{|p{0.9\columnwidth}|}
  \hline
  \textbf{CH4: Software architecture} \\ 
  \hline
\end{tabular}
\vspace{0.1cm}

AD requires a flexible and scalable architecture, which enables continuous updates~\cite{lee2024enhancing}. Therefore, its architecture is not limited to onboard computing; it also extends to cloud-based services that communicate with onboard systems through real-time communication~\cite{lee2024enhancing}.
Amalfitano et al. emphasize on the critical role of software architecture in complex safety-related applications such as AD~\cite{amalfitano2024characterizing}, a point also reported by Kugele et al. in ``Architecture as a Backbone for Safe DevOps in Automotive Systems''~\cite{kugele2022architecture}.

Since architecture does not evolve as quickly as software itself, it can become a bottleneck for functional growth, highlighting the need for continuous architecture along with evolving documentation and reasoning~\cite{bosch2022accelerating}.
Outdated architectural documentation in each development loop and the inability to assess cohesion and coupling between components are some of the reported issues~\cite{amalfitano2024characterizing}.
Additionally, due to centralized architecture and the possibility of both, safety-related and non-safety-related components coexisting on the same system, there is a risk of failure propagation, which could lead to the violation of safety requirements~\cite{kugele2022architecture}.


\emph{Sol4.1:} Ensuring Freedom From Interference (FFI) is essential for avoiding cascading failures in periodical updates~\cite{kugele2022architecture}.
To fulfill FFI and enable continuous development and integration after each innovation delivery, Kugele et al. propose hypervisors for partitioning between safety-related and non-safety-related software that is executed on the same processor~\cite{kugele2022architecture}. 
Kugele et al. also introduced \textit{containerisation} in their 2018 paper, ``Data-Centric Communication and Containerisation for Future Automotive Software Architectures''~\cite{16swcontain}. According to Lee et al. container technology on top of the operating system enables a division between safety and non-safety applications~\cite{lee2024enhancing}.

\emph{Discussion:} 
As a result of continuous function improvement or ODD expansion, new decomposition patterns need to be applied, which impose new requirements on the software architecture. For instance, in the case of Automotive Safety Integrity Level (ASIL) decomposition, the decomposed components must not have any common cause failures, which imposes requirements on shared resources. Hence, all these restrictions need to be considered, which may challenge flexibility.
Moreover, as the architecture extends to cloud-based services, which are not directly addressed in the mentioned safety standards, it becomes challenging to consider them in safety activities using conventional standards.

\emph{Industrial Context:} Nvidia highlights the importance of scalability in the SW architecture to enable functional growth, and support multiple levels of integrity and autonomy, from Level 2 to Level 5, through integration of additional hardware and software~\cite{NVIDIA-VSSA}. This also applies to different levels of safety, as OEMs may adjust their safety requirements (e.g., redundancy or diversity) in response to ODD or functionality expansions.

\vspace{0.3cm}
\begin{tabular}{|p{0.9\columnwidth}|}
  \hline
  \textbf{CH5: Verification and validation} \\ 
  \hline
\end{tabular}
\vspace{0.1cm}

Testing is essential for the certification and safety of autonomous systems; however, the infinite state-space makes it impractical to test all corner cases in a real-world setting~\cite{18siddique2020safetyops}$^{I}$, a challenge known as the ``long tail'' of events~\cite{hartmannsgruber2024improving}$^{I}$.
These events are critical as they are rare and they were most likely not part of the training data, leading to weaknesses in the software's ability to handle these events~\cite{john2023towards}.
Hence, it is essential to plan and design a granular verification and validation strategy for each technology used in AD, depending on their requirements, limitations, and weaknesses.
For instance, ML-based software testing involves three main phases, each with its own purpose: during model training, offline pre-deployment, and online post-deployment testing~\cite{borg2021aiq}. 
Adhering to such a thorough and lengthy testing process is challenging due to frequent changes in relatively short iterations~\cite{stapp2024istqb}. Moreover, modifying such a complex software requires extensive validation, significantly increasing development expenses~\cite{john2023towards}.
Ideally, the verification and validation aspects of safety argumentation should allow for minor system updates without requiring the repetition of major parts of the tests~\cite{hoss2022review}. However, this cannot be achieved without suitable methods, strategies, and technical approaches.

\emph{Sol5.1:} Simulations can cover a wider range of scenarios by using catalogues and randomising them~\cite{18siddique2020safetyops}$^{I}$.
It can be even used for Simulation-Aided Hazard and Risk Analysis (SAHARA), which assists a safety engineer in identifying of where to perform a Hazard Analysis Risk Assessment~\cite{meyers2019model}. However, three key aspects must be considered: Ensuring the relevance of the scenario to the ODD, maintaining traceability between the scenario and safety requirements, and calculating test coverage~\cite{18siddique2020safetyops}$^{I}$.
Software-in-the-Loop (SiL) is also proposed as a virtual environment for continuously testing new software by utilising the virtual version of the hardware and other software components~\cite{raghupatruni2021towards}$^{I}$, ~\cite{haas2024controlling}$^{I}$.
Model-based simulation~\cite{blanco2023comprehensive, mortstrand2023continuous}, ~\cite{tomar2024migration}$^{I}$, including software and relevant electrical and mechanical aspects~\cite{haas2024controlling}$^{I}$, can be employed for high-level testing and capturing design faults in the early phases of development.
Simulation models can also be used to generate synthetic raw data and its ground truth (e.g., object list) for testing perception algorithms without the need for physical sensors~\cite{hoss2022review} for diverse and unseen scenarios~\cite{schuster2023synthetic}$^{I}$ and data-driven validation~\cite{hartmannsgruber2024improving}$^{I}$, ~\cite{habibullah2024requirements}$^{I}$.
The synthetic and real-world data are then combined to avoid deviating from reality, addressing the so-called ``Reality Gap''~\cite{schuster2023synthetic}$^{I}$.

\emph{Sol5.2:} Before entering operative mode, where the software sends unsupervised commands to the actuators, the so called shadow mode approach allows the system to be validated by perceiving and planning without being executed~\cite{zeller2024toward}$^{I}$. Shadow mode tests new software versions under real operational conditions, measuring the readiness of it for real world implementation~\cite{blanco2023comprehensive}$^{I}$, ~\cite{hartmannsgruber2024improving}$^{I}$, ~\cite{25fayollas2020safeops}$^{I}$. 
This concept can be used to provide evidence for the safety case and achieve continuous learning and data collection in each version~\cite{22johansson2022continuous}$^{I}$, ~\cite{czarnecki2019software}. For instance, if discrepancies in detected objects occur between two decomposed paths during operation, the pertinent data is gathered to examine the underlying cause~\cite{22johansson2022continuous}$^{I}$.

\emph{Sol5.3:}~\cite{giaimo2019automotive, borg2021aiq}, ~\cite{25fayollas2020safeops}$^{I}$ discuss canary and A/B testing as methods for evaluating software during the operational phase, which is known as \textit{continuous experimentation}. Canary testing introduces new software to a random, small user sample to identify errors prior to full deployment, while A/B testing employs specific user selection criteria.

\emph{Sol5.4:} Test automation~\cite{stapp2024istqb, mortstrand2023continuous} improves efficiency and speed in each iteration, especially for tasks that are frequent, such as ML updates, which are continuously refined by collected data. As ML is the foundation of perception in AD, it is crucial to verify and validate its performance after each training or fine-tuning cycle in MLOps, or when the model is exposed to a new environment.
Employing techniques such as ``Gradient-weighted Class Activation Mapping (Grad-CAM)''~\cite{Borg2021Test} improves the explainability of the model and enables automation. Grad-CAM provides a heatmap that shows which part of the image the model focuses on to decide on the classification of an object, which can be automated by comparing the highlighted area with the expected one~\cite{Borg2021Test}.

\emph{Discussion:} Although simulation (Sol5.1) has benefits, including testing harmful scenarios without causing any hazard, correlating between simulation and reality is a challenging task~\cite{czarnecki2019software}, ~\cite{habibullah2024requirements}$^{I}$. Moreover, increasing the precision of simulations increases the computational cost of each run, making it impractical for achieving full coverage, especially for complex systems such as AD in multi-agent and complex environments. This highlights the need for smart selection of use cases of simulation and the necessity of bringing real world solutions such as shadow mode (Sol5.2) or continuous experimentation (Sol5.3) for safe use cases.
Test automation (Sol5.3) is beneficial as it reduces the risk of human errors due to its consistency and repeatability; however, it is not a replacement for ``human critical thinking''~\cite{stapp2024istqb}.

\emph{Industrial Context:}
To achieve rapid continuous integration, as highlighted in the Kodiak VSSA report~\cite{Kodiak-VSSA}, the latest build of each week is tested against the current stable software, and if the results show significant safety improvements, it is selected as the new stable version.
According to the Mercedes-Benz's VSSA report~\cite{mercedes2019introducing}, software components are tested using a well-defined test plan in various environments, such as Hardware-in-the-Loop (HIL) and Vehicle-in-the-Loop (VIL) tests conducted on proving grounds, before being tested on public roads.
According to the AutoX Team~\cite{AutoX-VSSA}, efficient validation tools such as simulation, HIL, and VIL enabled them to update the software within two days.
Minimizing dependency on public road testing can accelerate the software update process~\cite{IKE-VSSA}.
Since real-world testing has limitations, simulation enhances completeness by covering a wider range of scenarios, in which AD is tested~\cite{mercedes2019introducing}.
Hence, after a software update (e.g., for handling complex scenarios), testing in a simulation environment is conducted to ensure that the software still meets the requirements and that no significant regression has occurred~\cite{SAE_AVSC}.
For instance, Waymo's team, in multiple reports and publications~\cite{webb2020waymo, schwall2020waymo, Waymo-VSSA}, describes the use of simulation to evaluate AD's crash avoidance capability.
As another example, according to Aurora's VSSA report~\cite{aurora-Robotics-VSSA}, their software was tested for more than two million left turns in simulation before performing one in the real world. Rapid software development and bug fixing, without introducing any safety risks, are among the benefits of simulation~\cite{aurora-Robotics-VSSA}. Moreover, as testing all permutations of elements in a scenario is not feasible in the real world, simulation is a viable alternative~\cite{aurora-Robotics-VSSA}.
According to WAAbi's VSSA report~\cite{Waabi-VSSA}, a high-fidelity, closed-loop, end-to-end simulator (called Waabi World) is used for faster and more efficient testing rather than real-world testing.
In some cases, simulation is used to examine the performance of the AD software in an alternative scenario rather than relying solely on collected data~\cite{schwall2020waymo}.
For instance, if a test vehicle operator assesses a risk of collision and intervenes, then simulation can be used to examine the performance of the software without operator interference~\cite{schwall2020waymo}.
According to Waymo~\cite{webb2020waymo}, there is a need to rely on both simulation and real-world testing, as some tests are not possible in the real world and others are not possible in simulation.
Moreover, virtual test scenarios originate from data collected on public roads or closed courses; however, sometimes they are generated from scratch solely for the simulation environment~\cite{webb2020waymo}.
Another crucial source for scenario generation is by sourcing from crash databases~\cite{webb2020waymo}. Although crashes rarely occur, continuous improvement in the AD system's reactions during these scenarios may lead to either no collisions or a reduction in the severity of these events~\cite{webb2020waymo}.
Closed-course testing is also employed to validate the assumptions used in the development of simulation environments and to assess their accuracy~\cite{webb2020waymo}. For instance, the braking profile in simulation should replicate that of the actual vehicle~\cite{webb2020waymo}.
As safety rating organizations such as New Car Assessment Programs (NCAPs) focus on current products, it is also important to continuously monitor the development of test scenarios and metrics proposed and considered by them~\cite{webb2020waymo}.
The use of a combination of simulation and on-road testing after each software update is also described in other reports~\cite{TUSIMPLE-VSSA, Volkswagen-VSSA, ZOOX-VSSA, apple-VSSA, a2z-VSSA, Ford-VSSA}.

\vspace{0.3cm}
\begin{tabular}{|p{0.9\columnwidth}|}
  \hline
  \textbf{CH6: Operation phase} \\ 
  \hline
\end{tabular}
\vspace{0.1cm}

Autonomous systems operating in complex and unpredictable scenarios may encounter conditions that were not previously anticipated or even considered. This would lead to known unknowns (roughly predicted), unknown unknowns (completely missed) and difficulties in accurately quantifying the associated risks especially during design phase, according to Schleiss et al.~\cite{10consafetyad}. 
Moreover, model drift is considered a potential threat during the operational phase, highlighting the need to identify it and, if necessary, retrain and update the models~\cite{john2023towards}, ~\cite{Brajovic2023ModelRF}$^{I}$. This is why the continuous monitoring of released functions is required not only by automotive standards and regulations but also by broader frameworks such as the EU AI Act. Article 9 of the AI Act mandates the implementation of a risk management system for continuous risk assessment and evaluation during operation~\cite{Brajovic2023ModelRF}$^{I}$.
Additionally, an iterative process for evaluating and improving mitigation strategies is required to ensure that AI-enabled systems operate safely; otherwise, the system must be disabled (i.e., fail-safe)~\cite{Brajovic2023ModelRF}$^{I}$.

\emph{Sol6.1:} Runtime monitoring dynamically verifies the function against safety requirements and triggers relevant mechanisms in case of violations~\cite{furrer2022safety}. The runtime monitor receives the necessary internal and external information and checks the system's behavior against rules, contracts, or policies, which are often specified formally~\cite{furrer2022safety}. 
Gautham et al. implemented a model that enhances the hazard causation model of STPA to monitor the operation phase and detect hazards at run-time~\cite{STPAruntime}. The dynamic risk assessment framework is claimed to improve safety assurance by monitoring, evaluating and responding to risks in real time. This framework, together with Employing Micro-Operational Design Domains (µODDs), prevents having worst-case assumptions during the design phase~\cite{10consafetyad}. 
ML techniques, such as Out-of-Distribution (OOD) detection, can also be employed to identify inputs that ML models have not encountered before~\cite{hartmannsgruber2024improving}$^{I}$.
Additionally, f-divergence can be employed to measure the difference between the data distribution during the operational phase and the training data used in the development phase, helping to quantify data drift and detect significant deviations that may impact model performance~\cite{Brajovic2023ModelRF}$^{I}$.
OOD detection and f-divergence approaches are used not only to trigger relevant runtime risk mitigation strategies but also to collect and store data for model improvement.

\emph{Sol6.2:} According to regulations, it is mandatory to equip AD-capable vehicles by ``Data Storage System for Automated Driving (DSSAD)''~\cite{ALKS} to record and store critical events (e.g., detected collision, system failure). An automated system for rapidly identifying and reporting incidents, including near misses when human intervention prevents accidents, is proposed as a solution in~\cite{10consafetyad, ALKS}. Automation can encompass elements such as verifying assumptions and adjusting values, logging and reporting, narrowing down root cause search space, and distributing updates~\cite{10consafetyad}. For instance, utilising concepts such as safety cage, the DNN itself can employ novelty detection to continuously monitor the operational environment and report when uncertainty exceeds a certain limit~\cite{borg2018safely}$^{I}$.
The collected data will not only be used for offline development but can also be utilized for runtime learning and adaptation~\cite{Cerrolaza2024AI}$^{I}$. However, this concept, if applied without any constraints (e.g., limited actuation, limited adaptation), is not currently covered by existing standards~\cite{Cerrolaza2024AI}$^{I}$.

\emph{Discussion:}  Dynamic and Run-time Risk Assessment framework (Sol6.1) would try to prevent incidents or hazards to occur. If any incidents missed to be prevented, then smart triggers detect, record, and report (Sol6.2) them to the OEM, to prevent them and avoid potential accidents.
However, run-time monitoring imposes extra demands on hardware (e.g., processing and memory overhead), which must be balanced against hardware limitations~\cite{heyn2024empirical} to avoid the risk of hardware resource shortages.
ML performance assessment techniques, such as O\underline{O}D, can also be used to identify technological weaknesses of the components in the system and, if persistent, add new restrictions to the O\underline{D}D until those weaknesses are resolved.

\emph{Industrial Context:} 
As NHTSA asks all companies to report on their ``Data Recording'' for analyzing failures, malfunctions, or system degradation in the event of a crash~\cite{Torc-Robotics-VSSA}, the recorded data is used for event reconstruction and the evaluation of design and software performance, leading to continuous software improvement~\cite{GM2018self}.
This is especially crucial for perception and localization, which are responsible for reliable object and event detection as vital elements in ADS safety~\cite{mercedes2019introducing}. As challenging environmental situations or data mismatches between different sensors might lead to failures in perception or localization, these events are collected, analyzed, and used for continuous improvement~\cite{mercedes2019introducing}.

\vspace{0.3cm}
\begin{tabular}{|p{0.9\columnwidth}|}
  \hline
  \textbf{CH7: Safety argumentation} \\ 
  \hline
\end{tabular}
\vspace{0.1cm}

A \textit{safety case} demonstrates achievement of functional safety by providing a structured argument, supported by evidences derived from the prescribed activities in the product safety life-cycle. According to~\cite{myklebust2020agile}$^{I}$, it is called ``safety case'' since it must argue that the system is safe, similar to presenting a case in a court of law.

Safety cases of AD depend on factors such environmental conditions, which may shift during time, and system aspects, which would change on every update. This raises the need for continuously updating the safety case~\cite{18siddique2020safetyops}$^{I}$, otherwise it will be outdated and not valid~\cite{10consafetyad}.
Another reason for continuous update is Continues Integration and Deployment (CI/CD)~\cite{johanssonSEOOC}$^{I}$, which can be due to bug fixing~\cite{24safeopsbosch}$^{I}$, reconfiguration~\cite{10consafetyad} or changes in the system design~\cite{18siddique2020safetyops}$^{I}$.
Since safety is an overall system property, it is essential not only to prove the safety of the updated element but also to ensure that the safety of the entire system is maintained~\cite{hake2024integrating}$^{I}$.

The size and complexity of the system and its assurance case (e.g., safety case), as well as its change management, are the most significant problems raised by assurance case practitioners~\cite{Andrzej2023Development}.
According to Munk and Schweiser, a safety case update for a bug fix may take up to two weeks~\cite{24safeopsbosch}$^{I}$. This would take even longer for adding a feature or expanding the ODD, which would delay the feature update.
For instance, specifying ODD requires particular attention, as the safety case is based on the assumptions derived from ODD~\cite{weiss2024approach}$^{I}$; hence, its expansion would affect those assumptions and require updates to the safety argumentation.

\emph{Sol7.1:} Incremental safety assurance~\cite{hoss2022review}, agile safety case~\cite{myklebust2020agile}$^{I}$,
continuous safety case development~\cite{22johansson2022continuous}$^{I}$, ~\cite{Andrzej2023Development}, \cite{18siddique2020safetyops}$^{I}$ and the progressive release of a safety case by releasing each work product can prevent the need for a big bang approach, and leads to progressive functional safety assessments as recommended by ISO 26262~\cite{10consafetyad}. The planning for a safety case should begin at the start of the development to identify which evidence is needed 
 and how it can be generated during the development~\cite{myklebust2020agile}$^{I}$.
A well-connected framework between all components of a safety case, such as collected data, design and V\&V artefacts, is a prerequisite for a machine-readable, hierarchical, and continuous safety case~\cite{warg2024salience4cav}$^{I}$, \cite{18siddique2020safetyops}$^{I}$. Additionally, this approach leads to a forward-looking safety case as proposed by Warg et al.~\cite{20warg2019continuous}$^{I}$. This involves starting the safety case at the same time as development, allowing for the evaluation of safety arguments and strategies before implementation.

\emph{Sol7.2:} The presence of uncertainty during the operational phase of the system in complex and unpredictable environments necessitates dynamic safety assurance~\cite{sivakumar2024design}.
Dynamic safety case and run-time monitoring is an approach used to evaluate the level of confidence or uncertainty of a safety case~\cite{10consafetyad, sivakumar2024design}. 
The avionics industry employs this approach that starts with an assurance architecture, which presents the risk reduction due to function and components. Then, the assurance rational itself is developed and finally, the assurance confidence is quantified. Using this, the safety of the system is assured during run-time~\cite{10consafetyad}.

\emph{Sol7.3:} 
Modularisation of a safety case, which suggests that the safety case of the components can be developed separately as safety case fragments, is an effective approach to handle complexity~\cite{Andrzej2023Development}, ~\cite{20warg2019continuous}$^{I}$.
The safety case fragments are referenced by the overall safety case of the item to provide further details.
This approach is recommended also by ISO 26262 for distributed development~\cite{AliSEAASTPA}$^{I}$, ~\cite{20warg2019continuous}$^{I}$. This would reduce the needed re-certification effort of modules in case of reuse of them in the same context~\cite{Andrzej2023Development}, ~\cite{20warg2019continuous}$^{I}$.
Wardzinski and Jarzebowicz proposed the System Assurance Reference Model (SARM)~\cite{Andrzej2023Development}, a framework designed to facilitate assurance case development and management. This model allows for the decomposition of the assurance case and the automation of certain aspects of it.

\emph{Sol7.4:} ``Safety Element out of Context'' (SEooC)~\cite{ISO26262} enables suppliers to begin the component development in parallel with the OEM, meaning at the same time or even sooner. This approach prevents development delays caused by missing inputs from an OEM by assuming the needs of an OEM, which will be validated when such inputs are available. This approach allows the supplier to deliver solutions promptly once the OEM identifies a need~\cite{johanssonSEOOC}$^{I}$. Although there is a requirement to validate the supplier's assumptions and a risk of mismatches that may necessitate changes, both parties can always determine the most efficient adjustment in either the OEM's or supplier's context, minimizing costs and delays.

\emph{Sol7.5:} ``Proven in use'' and ``history in service'' argument can be applied to components that were not developed based on standards but have been used previously.
It relies on a monitoring system or component behaviour during operation, determining satisfactory confidence levels based on incident-free performance over time~\cite{25fayollas2020safeops}$^{I}$.

\emph{Sol7.6:} Quantitative or qualitative confidence estimation is useful to demonstrate the level of trust that can be placed on the safety case, key for continuously assuring safety~\cite{10consafetyad}, ~\cite{warg2024salience4cav}$^{I}$. Schleiss et al. gathered qualitative or quantitative approaches for estimating confidence~\cite{10consafetyad}. For instance, The ``Assurance Claim Points'' approach can be used to demonstrate that the system is safe, and a contrapositive approach to prove that the system is unsafe.
Safety Performance Indicators (SPIs) are quantitative metrics that use evidence and threshold comparison to make claims in safety cases, which help to measure claim satisfaction and detect incorrectness in safety cases~\cite{22johansson2022continuous}$^{I}$. SPIs also aid in the preparation of safety cases for future versions by continuously learning from the current version to be used in the next version of the safety case~\cite{22johansson2022continuous}$^{I}$.

\emph{Sol7.7:}
AI-based development assistance for tasks such as safety assessment and automated safety case construction can leverage Natural Language Processing (NLP) techniques, including LLMs~\cite{Cerrolaza2024AI}$^{I}$, ~\cite{sivakumar2024design}. For instance, in~\cite{sivakumar2024design, sivakumar2024prompting}, GPT-4 is used to generate safety cases in Goal Structuring Notation (GSN) format for an ML-based function and being compared with human generated safety case. 
According to the authors, ``in our rigorous evaluation, GPT-4 has displayed a remarkable level of competence, achieving an excellent grade of A in its ability to work with GSN~\cite{sivakumar2024prompting}.'' However, the author also highlights the importance of human oversight and review~\cite{sivakumar2024prompting}.

\emph{Sol7.8:} The contract-based design approach formalizes the interaction between different components and clarifies hidden assumptions by defining precise contracts~\cite{benveniste2018contracts, demi2024trustworthy}. This approach improves the system design process and enhances the distribution of responsibilities across stakeholders.
Safety and Security (SASE) contracts concept is an adaptation of contract-based design for ensuring safety and security of the system~\cite{Agirre2020Agile}$^{I}$. One improvement in SASE contracts is their dynamic and adaptable nature, allowing them to accommodate modifications in components' interactions, making them suitable for DevOps.
Combination of component-based designs and other approaches such as formal safety contracts can facilitate automation in safety case maintenance. However, since formal formulation is difficult, even semi-formal approaches offer benefits~\cite{20warg2019continuous}$^{I}$.
Moreover, the combination of modularisation and safety contracts allows for the item level safety case to only contain HARA and its validation, with a link to lower level safety cases for each module. This enables component reuse and use of modules in multiple products, as long as the contracts remain intact. Changes within the component that do not affect the safety contract will not impact an item's safety~\cite{20warg2019continuous}$^{I}$. Reusable patterns for arguments and contracts foster the understandability of the safety argument~\cite{20warg2019continuous}$^{I}$. 
According to~\cite{Modular2002Kelly}, modular architecture also lead to development costs reduction and improve maintainability.
It improves verification and validation by providing measurable criteria for system behavior that can be verified during both the development and operation phases~\cite{hake2024integrating}$^{I}$.

\emph{Discussion:} Since the safety case is the center of all safety activities, it can be considered as the most challenging aspect for DevOps. This is why it has been the focus of several publications as summarised before. Continuous Safety Case Development (Sol7.1) and Dynamic Safety Case (Sol7.2) emphasise that it is a living work product and not a one-time activity.
Fragmentation of safety case (Sol7.3) uses the divide-and-conquer approach to address system complexity and is an enabler for solutions such as SEooC (Sol7.4) and Proven in use (Sol7.5). 
Confidence estimation (Sol7.6) estimates uncertainty to avoid conservative restriction of ODD or specific solutions. Although confidence estimation is beneficiary, it is challenging to calculate it especially in an AD's dynamic environment.

\emph{Industrial Context:} VSSA reports are a publicly available snapshot of the safety argumentation of AD developers, which are discussed in previous challenges. As reported, companies mostly focus and report on verification and validation, while there is little coverage of design phase activities such as requirements engineering and safety analysis. According to Aurora’s Safety Case Framework~\cite{aurora-GSN, aurora-Robotics-VSSA}, which is presented in GSN, their framework continue to evoulve through their learnings from testing and expanding their operations.
Analyze, build, simulate and drive is presented as a loop leading to continuous improvement of the safety~\cite{GM2018self}. Incorporating learning and safety data into next versions of software would lead to safer designs~\cite{GM2018self}.
According to IKE's safety report~\cite{IKE-VSSA}, one of the biggest challenges of AD developers is ensuring safety during technology development due to rapid changes in hardware and software. Moreover, the complexity of system and its software which ``represent years of diligent engineering and safety
analysis, and in combination present even more complexity''~\cite{IKE-VSSA}.

\vspace{0.3cm}
\begin{tabular}{|p{0.9\columnwidth}|}
  \hline
  \textbf{CH8: Certification and assessment} \\ 
  \hline
\end{tabular}
\vspace{0.1cm}

Certification of continuous updates for vehicle software is challenging due to involvement of external stakeholders such as certification parties~\cite{kugele2022architecture}.
Traditional certification and compliance models face challenges in managing the dynamic nature of current software update practices in modern vehicles, not only for parts containing new software but also for previously certified functions that may be affected~\cite{santilli2023continuous}$^{I}$.
Normally, confirmation measures such as assessments are conducted when all activities are completed or near to the end, and this leads to long waiting time for review and feedback, which can delay updates. Moreover, changes in the software sometimes lead to the need for re-certification~\cite{Nouri2022Experience}$^{I}$.

\emph{Sol8.1:} Warg et al. propose that a combination of assessment and confirmation review with development activities can be used within the continuous development pipeline~\cite{20warg2019continuous}$^{I}$. This will also push organisations towards continuous execution of safety activities as suggested in some of the previously mentioned solutions.

\emph{Discussion:} Confirmation review, assessment, and certification can be partially automated as there are repetitive tasks such as checking for the existence of evidence and test coverage.
It is feasible since there are pre-existing checklists with defined criteria that can be implemented and checked automatically. \emph{Sol7.7} (LLM), \emph{Sol7.8} (contract-based), and \emph{Sol9.1} (automation) present techniques for this idea, which are discussed in detail.

\vspace{0.3cm}
\begin{tabular}{|p{0.9\columnwidth}|}
  \hline
  \textbf{CH9: Change impact analysis and tailoring} \\  
  \hline
\end{tabular}
\vspace{0.1cm}

The safety of a system is a property that results from its components and interactions, therefore, in each feature release, the entire system, including the hardware, must be analysed for the impact of changes, even if the change is only on the software. Incorporating this into rapid iteration cycles may not be seen feasible without compromising either the rapidity or the completeness of the required tasks. The latter case is the reason why the safety community does not fully welcome DevOps~\cite{24safeopsbosch}$^{I}$. For instance, if a new feature is added or changes are made to any existing feature, it is necessary to analyse the impact on hazard analysis and risk assessment for both ISO 26262 and ISO 21448~\cite{24safeopsbosch}$^{I}$. If a new hazard is identified, at least one safety goal or validation target and acceptance criterion should be assigned to it. This process should be continued towards the most granular abstraction level in the functional and technical safety concept~\cite{24safeopsbosch}$^{I}$, ~\cite{Nouri2022Experience}$^{I}$.

\emph{Sol9.1:} Incorporating automation into the continuous safety case process as presented in \emph{Sol7.1} enables quick assessment of the impact of changes in component activities on the overall safety case~\cite{18siddique2020safetyops}$^{I}$. Automating repetitive tasks can also improve efficiency and allow safety engineers to focus on intellectual activities~\cite{25fayollas2020safeops}$^{I}$.
Automation of establishing traceability between artifacts and performing change impact analysis is challenging due to diversity of artifacts and their format (e..g., requirements in natural language, design artifacts in UML, and C++ codes).
However, automation can be achieved using techniques such as artifacts using XML-based representation, Natural Language Processing (NLP), and matching algorithms~\cite{Rubasinghe2018Traceability}.
``Policy as code''~\cite{santilli2023continuous}$^{I}$ is an approach for automating compliance checks against requirements from standards and regulations throughout the DevOps cycles.

\emph{Sol9.2:} Utilising Model-Based Systems Engineering (MBSE) and integrating safety activities therein can enhance not only efficiency but also synchronisation between development and safety activities~\cite{24safeopsbosch}$^{I}$, ~\cite{czarnecki2019software}, ~\cite{Till2024Application}$^{I}$, ~\cite{tomar2024migration}$^{I}$. The aSET project proposes a formal safety model to improve the iterative design process using MBSE~\cite{meyers2019model}.

\emph{Sol9.3:} 
Traceability between artifacts, from requirements to field monitoring, is crucial for tracking how changes in one artifact impact others in each iteration~\cite{24safeopsbosch}$^{I}$~\cite{Batot2021Traceability}$^{I}$~\cite{Rubasinghe2018Traceability, Andrzej2023Development}~\cite{santilli2023continuous}$^{I}$. 
Traceability models~\cite{24safeopsbosch}$^{I}$~\cite{Batot2021Traceability}$^{I}$ are enablers for ensuring consistent and comprehensive linking of safety artifacts in the development lifecycle. For instance, updating the Traceability Information Model (TIM) during each iteration, such as when there are changes in the safety goals or its attributes, can automatically identify the affected safety requirements in the Functional Safety Concept (FSC) and the Technical Safety Concept (TSC). This is then extended to include safety analysis, verification, validation, and confirmation measures such as assessments~\cite{24safeopsbosch}$^{I}$. 
Traceability graphs can also be employed to improve visualisation and identify the impact of changes on each artifact (node) by tracking the traceability links (edges) between them~\cite{Rubasinghe2018Traceability}, thereby enabling automation (\emph{Sol9.1}).
Carlan et al. proposed a semi-automated approach for analysing the impact of changes in safety arguments by enhancing the traces between argumentation elements and models of assurance artefacts~\cite{32simonCIA}$^{I}$. The argument and its elements are implemented in a model-based environment, and by monitoring the changes in the specified ODD, the impact of these change on the artefacts has been identified.
Blockchain-enabled frameworks such as the one proposed by~\cite{demi2024trustworthy} improve traceability of changes in software artifacts even in decentralized distributed development.

\emph{Discussion:} Although automation is proposed as a solution in \textit{Sol9.1} and recommended in \textit{CH8}, it is important that the outputs of automated solutions still need to be reviewed by engineers to ensure accuracy and comprehensiveness.

\emph{Industrial Context:}
According to \cite{schnelle2023ads}, the impact of software updates on the safety assessment and evaluation process in safety cases is among the open research challenges. Moreover, \cite{schnelle2023ads} questioned the practicality of the ``Change Anything, Change Everything'' (CACE) approach, proposed by~\cite{sculley2015hidden}.

\vspace{0.3cm}
\begin{tabular}{|p{0.9\columnwidth}|}
  \hline
  \textbf{CH10: Tooling} \\  
  \hline
\end{tabular}
\vspace{0.1cm}

Recently, there has been a shift in using more advanced software tools in system and software development. 
Additionally, as mentioned earlier, most of the solutions involve automation, which in turn requires a dedicated tool chain for each solution to make it automated.
For instance, continuous Integration and Deployment
(CI/CD) are enablers for preserving the safety and quality of software while reducing the time between development and operations~\cite{borg2021aiq, fuchs2024validation}.
The concepts of CI/CD pipelines and version control systems along with their associated tools are commonly utilised in DevOps ecosystems, including automotive industry~\cite{24safeopsbosch}$^{I}$.
Nevertheless, tools used in safety activities are not as advanced as the tools used for visualisation and code reviewing, which has led to safety-related activities lagging behind~\cite{18siddique2020safetyops}$^{I}$. An example of this are visualisation features, which safety-related tools often lack and are not easy to integrate with existing tools~\cite{18siddique2020safetyops}$^{I}$. 
Moreover, limited tool support and insufficient integration of the assurance case development process with other tools involved in development are cited as root causes of the challenges practitioners face during the development of assurance cases~\cite{Andrzej2023Development}.

\emph{Sol10.1:} Open APIs facilitate the connection of different safety tools, which enables data-driven development for safety activities. For instance, it enables data exchange between safety analysis methods and even with simulation or real-world verification and validation tools~\cite{18siddique2020safetyops}$^{I}$. Establishing traceability between the results and assumptions in safety analysis methods, simulation, and real-world testing can enable cross-validation, avoiding mismatches and improving impact analysis. Furthermore, by allowing self-updates for safety analysis and simulation tools when real-world data changes, we can ensure their results remain accurate and up-to-date~\cite{18siddique2020safetyops}$^{I}$.

\emph{Sol10.2:} Modernising software safety tools by adding tools to summarise the results of automatic impact analysis, e.g. by improving visualisation features~\cite{18siddique2020safetyops}$^{I}$ or implementing a safety management dashboard~\cite{24safeopsbosch}$^{I}$ is a desirable solution. Another possibility for addressing \emph{CH10} is to integrate safety activities into already existing software development tools. An example of this approach is the concept of ``doc as code''~\cite{18siddique2020safetyops}$^{I}$.

\emph{Discussion:} According to ISO 26262, ``Confidence in the use of software tools'' is mandatory in safety-related tool chains~\cite{24safeopsbosch}$^{I}$. This process begins by identifying the Tool Confidence Level (TCL), and then, a suitable method should be chosen based on the ASIL of the function~\cite{ISO26262, fuchs2024validation}. If a tool has an impact on ASIL requirements and its failure is not detectable, then the tool validation or development process itself must comply with ISO 26262~\cite{ISO26262, fuchs2024validation}, which can be time-consuming and that requires significant effort. Hence, it is challenging to adopt innovative and user-friendly tools for safety-related applications.

\emph{Industrial Context:}
Developing and integrating internal tools such as Requirement Tracking, Configuration Management, and Verification \& Validation tools enables a robust, rapid, and iterative safety ecosystem~\cite{IKE-VSSA}. For instance, if the value in a requirement is updated, the software is automatically tested against this new value~\cite{IKE-VSSA}.

\vspace{0.3cm}
\begin{tabular}{|p{0.9\columnwidth}|}
  \hline
  \textbf{CH11: Change in one domain affects others} \\  
  \hline
\end{tabular}
\vspace{0.1cm}

As pointed out by Nouri et al., technical solutions of functional safety, SOTIF, and cyber-security are interconnected and can sometimes contradict each other~\cite{Nouri2022Experience}$^{I}$. This adds another complexity dimension to the already mentioned challenges.

\emph{Sol11.1:} The UP2DATE project~\cite{Agirre2020Agile}$^{I}$ proposed safety-security co-engineering, and the SASE contracts concept as described in \emph{Sol7.8}, which considers both safety and cyber-security simultaneously.

\emph{Discussion:} 
Seperation of activities in each domain might be one root cause of this challenge, which is avoided in~\cite{Agirre2020Agile}$^{I}$. Mapping of acitivities in each domain to the equivalent activity in other domains might lead to better allocation of activities in the organization, which reduce the misalignment between domains. Integration of activities also might be a solution. For instance, integration of HARA in the context of ISO 26262~\cite{ISO26262}, Hazard Identification \& Evaluation in ISO 21448~\cite{sotif}, and Threat Analysis and Risk Assessment (TARA) in ISO 21434~\cite{isocs} can be a remedy for avoiding misalignment from the very first activities in all three domains.
Moreover, some solutions identified for other challenges could potentially be employed to this challenge. For example, expanding \emph{Sol9.3} (i.e. TIM) between disciplines could help to identify the direct and indirect impact of changes across teams and disciplines~\cite{24safeopsbosch}$^{I}$.

\emph{Industrial Context:} One of the key safety design elements is the cybersecurity process of AD developers, as recommended for reporting by NHTSA~\cite{NHTSA2017VisionForSafety}. All VSSA reports describe their cybersecurity process and their continuous monitoring and improvement efforts to safeguard their AD systems against cybersecurity threats and attacks, such as the one reported in~\cite{marsauto-VSSA}.

\renewcommand\arraystretch{1}

\begin{sidewaystable*}
    \centering
    \caption{Mapping challenges (RQ1) to solutions (RQ2) proposed in literature (part 1) along with their supporting references.\\ Papers with industrial authors are indicated by [X]$^{I}$, and the total number is reported in column `I'.}
    \label{tab:solutions1}

    \begin{tabular}{|p{40mm}p{1mm}|p{120mm}p{40mm}|p{1mm}|}
        \hline
        \textbf{CHX \& CH. Ref.} & &\textbf{Justification of mapping} & \textbf{Sol. Ref.} & \textbf{I}\\
        \hline
        \hline
CH1: Safety and software workflow separation & & Sol1.1: In addition to all independent safety experts, a ``Quality Assurer'' with relevant and proper safety knowledge should be added to the team to ensure the quality of safety activities. &  ~\cite{hanssen2018safescrum}$^{I}$, ~\cite{myklebust2020agile}$^{I}$ & 2\\
~\cite{18siddique2020safetyops}$^{I}$ \cite{20warg2019continuous}$^{I}$~\cite{25fayollas2020safeops}$^{I}$& & & & \\
\hline
CH2:Requirement updates& &Sol2.1: Data-driven and AI-driven development besides traditional requirements engineering methods. & \cite{1jantango, kugele2022architecture, 14KrystofREAD}, ~\cite{20warg2019continuous}$^{I}$, ~\cite{habibullah2024requirements}$^{I}$, ~\cite{warg2024salience4cav}$^{I}$ & 3\\
 ~\cite{14KrystofREAD, 1jantango} ~\cite{Nouri2022Experience}$^{I}$ ~\cite{habibullah2024requirements}$^{I}$ & & & & \\
\hline
CH3: Safety analysis methods & &Sol3.1: Integrated safety analysis framework with live connection to system design and field data. & ~\cite{18siddique2020safetyops}$^{I}$, ~\cite{thomas2023toward}, ~\cite{AliSEAASTPA}$^{I}$ & 2\\
\cline{3-5}
~\cite{18siddique2020safetyops}$^{I}$, ~\cite{thomas2023toward} ~\cite{AliSEAASTPA}$^{I}$& & Sol3.2: Natural Language Processing (NLP) techniques, such as BERT and GPT, can be used to improve the speed and efficiency of safety analysis. & \cite{baumler2024predicting}, ~\cite{zeller2024toward}$^{I}$ ~\cite{nouri2024RELLM}$^{I}$, ~\cite{Nouri2024CAINLLM}$^{I}$ & 3\\
\hline
CH4: Software architecture & &Sol4.1: Partitioning by hypervisors for freedom from interference and employing containerisation. & \cite{kugele2022architecture, 16swcontain, lee2024enhancing} & - \\
~\cite{lee2024enhancing, amalfitano2024characterizing, kugele2022architecture, bosch2022accelerating} & & & & \\
\hline
CH5: Verification and validation & &Sol5.1: Simulation can cover a wider range of relevant scenarios, trace results back to requirements and enable synthetic data generation for data driven verification. & ~\cite{18siddique2020safetyops}$^{I}$, \cite{meyers2019model, hoss2022review, mortstrand2023continuous}, ~\cite{hartmannsgruber2024improving}$^{I}$, ~\cite{schuster2023synthetic}$^{I}$, ~\cite{raghupatruni2021towards}$^{I}$, ~\cite{habibullah2024requirements}$^{I}$ ~\cite{blanco2023comprehensive}$^{I}$, ~\cite{tomar2024migration}$^{I}$, ~\cite{haas2024controlling}$^{I}$ & 8\\
\cline{3-5}
~\cite{18siddique2020safetyops}$^{I}$, \cite{borg2021aiq, stapp2024istqb, hoss2022review, john2023towards}, ~\cite{hartmannsgruber2024improving}$^{I}$ & &Sol5.2: Shadow mode testing can continuously verify safety and performance requirements. & ~\cite{22johansson2022continuous}$^{I}$, \cite{czarnecki2019software}, ~\cite{hartmannsgruber2024improving}$^{I}$, ~\cite{zeller2024toward}$^{I}$, ~\cite{25fayollas2020safeops}$^{I}$, ~\cite{blanco2023comprehensive}$^{I}$ & 5\\
\cline{3-5}
 & &Sol5.3: Continuous experimentation (canary and A/B) during the operational phase. & \cite{giaimo2019automotive, borg2021aiq}, ~\cite{25fayollas2020safeops}$^{I}$  & 1\\
\cline{3-5}
 & &Sol5.4: Test automation can be achieved by employing relevant techniques for each technology, such as ``Gradient-weighted Class Activation Mapping (Grad-CAM)'' for ML components. & \cite{stapp2024istqb, mortstrand2023continuous, Borg2021Test} & -\\
\hline

        \hline

    \end{tabular}

\end{sidewaystable*}

\begin{sidewaystable*}
    \centering
    \caption{Mapping challenges (RQ1) to solutions (RQ2) proposed in literature (part 2) along with their supporting references.\\ Papers with industrial authors are indicated by [X]$^{I}$, and the total number is reported in column `I'.}
    \label{tab:solutions2}

    \begin{tabular}{|p{40mm}p{1mm}|p{120mm}p{40mm}|p{1mm}|}
        \hline
        \textbf{CHX \& CH. Ref.} & &\textbf{Justification of mapping} & \textbf{Sol. Ref.} & \textbf{I}\\
        \hline
        \hline

CH6: Operation phase & &Sol6.1: Dynamic and run-time risk assessment framework, along with ML techniques such as Out-of-Distribution (OOD) detection, can identify risk, weaknesses and limitations of the system during the operation phase. & \cite{10consafetyad,STPAruntime, furrer2022safety} ~\cite{Brajovic2023ModelRF}$^{I}$ ~\cite{hartmannsgruber2024improving}$^{I}$ & 2\\
\cline{3-5}
~\cite{10consafetyad, john2023towards} ~\cite{Brajovic2023ModelRF}$^{I}$ & &Sol6.2: Operation phase monitoring by identifying relevant triggers and incident reporting & \cite{ALKS}$^{I}$ \cite{10consafetyad} ~\cite{Cerrolaza2024AI}$^{I}$ ~\cite{borg2018safely}$^{I}$ & 3
\\
        \hline

        CH7: Safety argumentation & &Sol7.1: Progressive functional safety assessments and continuous development and release of safety case & \cite{hoss2022review, Andrzej2023Development, 10consafetyad}~\cite{20warg2019continuous}$^{I}$, ~\cite{18siddique2020safetyops}$^{I}$, ~\cite{myklebust2020agile}$^{I}$, ~\cite{warg2024salience4cav}$^{I}$, ~\cite{22johansson2022continuous}$^{I}$& 5\\
\cline{3-5}
 ~\cite{18siddique2020safetyops}$^{I}$, ~\cite{johanssonSEOOC}$^{I}$ ~\cite{24safeopsbosch}$^{I}$, \cite{10consafetyad, Andrzej2023Development} ~\cite{weiss2024approach}$^{I}$, ~\cite{myklebust2020agile}$^{I}$, ~\cite{hake2024integrating}$^{I}$& &Sol7.2: Dynamic safety cases and run-time monitoring to assure the system safety during run-time & \cite{sivakumar2024design, 10consafetyad} & - \\
\cline{3-5}
 & &Sol7.3: Fragmentation of safety cases and safety contracts to reduce re-certification efforts & \cite{Andrzej2023Development} ~\cite{AliSEAASTPA}$^{I}$ ~\cite{20warg2019continuous}$^{I}$, ~\cite{25fayollas2020safeops}$^{I}$& 3\\
\cline{3-5}
 & &Sol7.4: Safety Element out of Context for component development teams in distributed development & 
~\cite{johanssonSEOOC}$^{I}$ & 1\\
\cline{3-5}
 & &Sol7.5: ``Proven in use'' argument for components not compliant with standards but incident-free & ~\cite{25fayollas2020safeops}$^{I}$  & 1\\
\cline{3-5}
 & &Sol7.6: Confidence estimation to prevent excessive conservatism in safety limitations & \cite{10consafetyad}, ~\cite{22johansson2022continuous}$^{I}$, ~\cite{warg2024salience4cav}$^{I}$& 2\\
\cline{3-5}
 & & Sol7.7: AI-based development assistance leverages Large Language Models for tasks such as safety assessment and automated safety case construction. & ~\cite{Cerrolaza2024AI}$^{I}$, \cite{sivakumar2024design, sivakumar2024prompting} & 1\\
\cline{3-5}
 & & Sol7.8: Formalising the interaction between different components through contract-based design.  & \cite{benveniste2018contracts, demi2024trustworthy, Modular2002Kelly}, ~\cite{Agirre2020Agile}$^{I}$, ~\cite{hake2024integrating}$^{I}$, ~\cite{20warg2019continuous}$^{I}$ & 3\\
\hline

    \end{tabular}

\end{sidewaystable*}

\begin{sidewaystable*}

    \centering
    \caption{Mapping challenges (RQ1) to solutions (RQ2) proposed in literature (part 3) along with their supporting references.\\Papers with industrial authors are indicated by [X]$^{I}$, and the total number is reported in column `I'.}
    \label{tab:solutions3}

    \begin{tabular}{|p{40mm}p{1mm}|p{120mm}p{40mm}|p{3mm}|}
        \hline
        \textbf{CHX \& CH. Ref.} & &\textbf{Justification of mapping} & \textbf{Sol. Ref.} & \textbf{I}\\
        \hline
        \hline

CH8: Certification and assessment & &Sol8.1: Incorporating assessment and confirmation review with development activities & ~\cite{20warg2019continuous}$^{I}$ & 1\\
~\cite{santilli2023continuous}$^{I}$, ~\cite{Nouri2022Experience}$^{I}$ & & & &\\
\hline
CH9: Change impact analysis and tailoring & &Sol9.1: Automation in the safety case process and assessment & \cite{Rubasinghe2018Traceability} ~\cite{santilli2023continuous}$^{I}$, ~\cite{18siddique2020safetyops}$^{I}$, ~\cite{25fayollas2020safeops}$^{I}$  & 3\\
\cline{3-5}
 ~\cite{24safeopsbosch}$^{I}$, ~\cite{Nouri2022Experience}$^{I}$ & &Sol9.2: Model-Based System Engineering as an enabler for having machine readable design & ~\cite{24safeopsbosch}$^{I}$, \cite{czarnecki2019software,meyers2019model}, ~\cite{Till2024Application}$^{I}$, ~\cite{tomar2024migration}$^{I}$& 3\\
\cline{3-5}
 & &Sol9.3: Traceability Information Model and graph for design aspects and safety artefacts and block-chain & ~\cite{24safeopsbosch}$^{I}$, ~\cite{32simonCIA}$^{I}$, ~\cite{Batot2021Traceability}$^{I}$, \cite{Rubasinghe2018Traceability, demi2024trustworthy, Andrzej2023Development}, ~\cite{santilli2023continuous}$^{I}$ & 4\\
\hline
CH10: Tooling & & Sol10.1: Open APIs to facilitate data exchange between safety analysis, simulation, and V\&V & ~\cite{18siddique2020safetyops}$^{I}$ & 1\\
\cline{3-5}
~\cite{24safeopsbosch}$^{I}$, \cite{borg2021aiq, fuchs2024validation, Andrzej2023Development}, ~\cite{18siddique2020safetyops}$^{I}$ & &Sol10.2: Modernising safety tools and integrating safety activities into software development tools & ~\cite{24safeopsbosch}$^{I}$, ~\cite{18siddique2020safetyops}$^{I}$& 2\\
\hline
CH11: Change in one domain affects others & &Sol11.1: safety-security co-engineering and SASE contracts concept, enable simultaneous multi-domain considerations. &  ~\cite{Agirre2020Agile}$^{I}$ & 1\\
~\cite{Nouri2022Experience}$^{I}$ & & & &\\

        \hline

    \end{tabular}

\end{sidewaystable*}

\section{Conclusion} \label{sec:discussion}

Our systematic analysis and synthesis of the 319 extracted studies show that there are 11 clusters of challenges (RQ1) for using DevOps for safety-related systems such as AD. The identified challenges are mapped to the essential safety activities in the DevSafeOps loop and presented in Figure~\ref{fig:DevOps}.
Potential solutions and relevant DevOps safety (i.e., DevSafeOps) activities are mapped to each challenge (RQ2). The findings are presented in Section~\ref{sec:results} and Table~\ref{tab:solutions1}, ~\ref{tab:solutions2} and ~\ref{tab:solutions3}, which aggregates and represents the complex mapping between existing challenges and mitigation strategies to them.

As discussed in Section~\ref{sec:results}, we have identified gaps in each challenge when considering the proposed solutions (RQ3), such as the prerequisites needed for certain solutions and the potential to apply and study some solutions for other challenges.
Additional investigation is also required to assess the coverage and effectiveness of solutions for each challenge.
For instance, requirements engineering for DevOps is deemed ``underexplored'' by Czarnecki~\cite{14KrystofREAD}.
In addition to all aforementioned challenges, DevOps could be extended to hardware aspects of sensors, controllers and actuators~\cite{thomas2023toward, kugele2022architecture}, ~\cite{Nouri2022Experience}$^{I}$.

As shown in Table~\ref{tab:solutions1}, ~\ref{tab:solutions2} and ~\ref{tab:solutions3}, some publications provide a high-level overview, and discuss multiple challenges and corresponding solutions, such as ~\cite{ czarnecki2019software,kugele2022architecture,10consafetyad,meyers2019model}, ~\cite{22johansson2022continuous}$^{I}$, ~\cite{24safeopsbosch}$^{I}$, ~\cite{20warg2019continuous}$^{I}$, \cite{18siddique2020safetyops}$^{I}$, ~\cite{25fayollas2020safeops}$^{I}$. The rest proposed one or multiple solutions for one challenge such as~\cite{1jantango,14KrystofREAD,16swcontain, giaimo2019automotive}, ~\cite{borg2018safely}$^{I}$, ~\cite{raghupatruni2021towards}$^{I}$.
Some solutions were discussed or investigated by multiple references, but no inconsistencies were observed. For instance, no identified reference questioned the validity or relevance of any of the identified challenges, or solutions proposed.

Certain solutions identified in this study, such as contract-based design (\emph{Sol7.8}) and Large Language Models (\emph{Sol7.7} and \emph{Sol3.2}), may be applicable to other challenges. However, they were not explicitly proposed by the reviewed publications. This highlights the need for further research in this area to explore potential cross-cutting solutions.
Moreover, Some solutions such as Model-Based System Engineering (\emph{Sol9.2})~\cite{24safeopsbosch}$^{I}$, ~\cite{czarnecki2019software}) are prerequisite for other solutions such as Model-based simulation (\emph{Sol5.1}) ~\cite{mortstrand2023continuous}, ~\cite{blanco2023comprehensive}$^{I}$, ~\cite{tomar2024migration}$^{I}$, meaning that if the foundational solutions are not implemented, the dependent solutions cannot be applied effectively.

During our study, we observed a limited number of peer-reviewed publications that address challenges and potential solutions for ensuring safety in DevOps, particularly in the context of autonomous driving. Hence, to ensure a comprehensive review, we did not exclude relevant non-peer-reviewed papers, selecting two papers~\cite{warg2024salience4cav}$^{I}$ ~\cite{18siddique2020safetyops}$^{I}$ that was deemed appropriate meeting the criteria from our review protocol. However, it should be noted that challenges and proposed solutions in these references may require validation, despite being valuable contributions written by a practitioner. As such, the proposed challenges and solutions represent potential topics for future studies.

Integrating new technologies into safety activities is crucial for improving safety of AD and development speed, while being careful not to over-promise. As most documents are in natural text, LLMs (\emph{Sol7.7} and \emph{Sol3.2}) can be employed in various aspects, including requirements engineering, safety analysis, testing, and change impact analysis~\cite{Rubasinghe2018Traceability}.
Moreover, due to their capability to accept inputs in the form of code, their application is not limited to tasks involving natural language.
However, their capabilities, weaknesses, and limitations need to be carefully studied and paired with quality assurance to ensure the validity of the results.

Solutions such as \emph{Sol9.1}, \emph{Sol6.2}, and \emph{Sol3.3} are improving the efficiency and speed by using automation. As discussed in CH10 (tooling) and highlighted by ~\cite{Andrzej2023Development, sivakumar2024design}, ~\cite{zeller2024toward}$^{I}$ ~\cite{nouri2024RELLM}$^{I}$, developers shall be in the loop to review and confirm the output of these systems. The weaknesses and limitations of such systems should be identified, and relevant avoidance or mitigation strategies should be designed before fully relying on these systems in safety-related domains such as AD.

When adopting DevOps, potential dilemmas may arise such as selecting speed at the expense of other criteria like quality, which is not viable for safety-related applications. In AD, while safety is an absolute necessity, the speed would be sacrificed due to the identified challenges. Employing identified solutions can increase the speed in DevSafeOps, satisfying both safety and rapidity concurrently.

Future work, in addition to addressing the major gaps, should explore the possibility of using the proposed solutions for other challenges. Furthermore, taking a grounded theory approach to investigate and validate the data used from grey literature in this study would be a natural next step.

\section*{Acknowledgments}
This work has been partially supported by Sweden’s Innovation Agency (Vinnova, diarienummer: 2021-02585), and by the Wallenberg AI Autonomous Systems and Software Program (WASP) funded by the Knut and Alice Wallenberg Foundation.

\section*{Disclaimer}
The views and opinions expressed are those of the authors and do not necessarily reflect the official policy or position of Volvo Cars.

\section*{The use of generative AI and AI-assisted technologies}
During the preparation of this work the first author used ChatGPT in order to check the grammar and spelling to improve readability. After using this tool, the authors reviewed and edited the content as needed and take full responsibility for the content of the publication.

\end{document}